\documentclass[11pt]{article}
\usepackage[latin1]{inputenc}
\usepackage{amsmath}
\usepackage{amsfonts}
\usepackage{amssymb}
\usepackage{latexsym}
\usepackage{graphicx}
\usepackage[a4paper]{geometry}
\usepackage{color}

\newtheorem{theorem}{Theorem}
\newtheorem{proposition}{Proposition}
\newtheorem{conjecture}{Conjecture}
\newtheorem{lemma}{Lemma}

\newtheorem{corollary}{Corollary}

\newenvironment{proof}{\noindent \emph{Proof.}\ }{\hfill$\Box$\vspace{1em}}

\begin{document}

\title{Asteroids in rooted and directed path graphs\footnote{Research
  supported by the Natural Sciences and Engineering Research Council
  of Canada (NSERC)}}

\author{ Kathie Cameron\thanks{Department of Mathematics, Wilfrid
    Laurier University, Waterloo, Ontario, Canada, N2L 3C5. \texttt{email: kcameron@wlu.ca}}\ \ \
  \
Ch\'inh T. Hoàng\thanks{Department of Physics and Computer Science,
  Wilfrid Laurier University, Waterloo, Ontario, Canada, N2L 3C5. \texttt{email: choang@wlu.ca}}\
\ \ \
Benjamin L{\'e}v{\^e}que\thanks{Department of Physics and Computer
  Science, Wilfrid Laurier University, Waterloo, Ontario, Canada, N2L 3C5. \texttt{email:
    bleveque@wlu.ca}}\ \ \ \ }
\maketitle

\begin{abstract}
An asteroidal triple is a stable set of three vertices such that
each pair is connected by a path avoiding the neighborhood of the
third vertex. Asteroidal triples play a central role in a
classical characterization of interval graphs by Lekkerkerker and
Boland. Their result says that a chordal graph is an interval graph if and
only if it contains no asteroidal triple. In this paper, we prove
an analogous theorem for directed path graphs which are the
intersection graphs of directed paths in a directed tree. For this
purpose, we introduce the notion of a strong path. Two non-adjacent
vertices are linked by a strong path if either they have
a common neighbor or they are the endpoints of two vertex-disjoint
chordless paths satisfying certain conditions. A strong
asteroidal triple is an asteroidal triple such that
each pair is linked by a strong path. We prove that a chordal graph is a directed
path graph if and only if it contains no strong asteroidal triple.
We also introduce a related notion of asteroidal quadruple, and conjecture a
characterization of rooted path graphs which are the intersection
graphs of directed paths in a rooted tree.
\end{abstract}

\section{Introduction}\label{sec:intro}

A \emph{hole} is a chordless cycle of length at least four.  A graph
is a \emph{chordal graph}  if it contains no hole as an induced
subgraph.  Gavril~\cite{Gav74} proved that a graph is chordal if and
only if it is the intersection graph of a family of subtrees of a
tree.  In this paper, whenever we talk about the intersection of
subgraphs of a graph we mean that the \emph{vertex sets} of the
subgraphs intersect.

\begin{figure}[h]
  \centering
\includegraphics[scale=1]{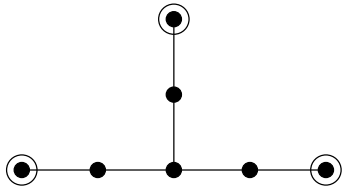}
  \caption{Graph containing an asteroidal triple}
  \label{fig:asteroidaltriple}
\end{figure}

A graph is an \emph{interval graph} if it is the intersection
graph of a family of intervals on the real line; or equivalently,
the intersection graph of a family of subpaths of a path. An
\emph{asteroidal triple} in a graph $G$ is a set of three non-adjacent vertices such that for any two of them, there exists a
path between them in $G$ that does not intersect the neighborhood
of the third. The graph of Figure~\ref{fig:asteroidaltriple} is an
example of a graph that minimally contains an asteroidal triple;
the three vertices forming the asteroidal triple are circled.

The following classical theorem was proved by Lekkerkerker and
Boland.
\begin{theorem}[\cite{LB62}]\label{th:intervalgraph}
  A chordal graph is an interval graph if and only if it contains
  no  asteroidal triple.
\end{theorem}

Lekkerkerker and Boland~\cite{LB62} derived from
Theorem~\ref{th:intervalgraph} the list of minimal forbidden
subgraphs for interval graphs (see Figure~\ref{fig:interval}).

The class of path graphs lies between interval graphs and
chordal graphs. A graph is a \emph{path graph} if it is the
intersection graph of a family of subpaths of a tree. Lévêque,
Maffray and Preissman~\cite{LMP08} found a characterization of path graphs by
forbidden subgraphs (see Figure~\ref{fig:path}).

Two variants of path graphs have been defined when the tree is a
directed graph. A \emph{directed tree} is a directed graph whose
underlying undirected graph is a tree. A graph is a \emph{directed
  path graph} if it is the intersection graph of a family of directed
subpaths of a directed tree. Panda~\cite{Pan99} found a
characterization of directed path graphs by forbidden subgraphs (see
Figure~\ref{fig:directed}).  A \emph{rooted tree} is a directed tree
in which the path from a particular vertex $r$ to every other vertex
is a directed path; vertex $r$ is called the \emph{root}. A graph is a
\emph{rooted path graph} if it is the intersection graph of a family
of directed subpaths of a rooted tree. The problem of finding a
characterization of rooted path graphs by forbidden subgraphs is still
open.

Clearly, we have the following inclusions between the classes considered :

\begin{quote}
\begin{center}
interval $\subset$ rooted path $\subset$ directed path $\subset$ path  $\subset$ chordal
\end{center}
\end{quote}
In this paper, we study directed path graphs and rooted path
graphs. Our main result is a characterization of directed path
graphs analogous to the theorem of Lekkerkerker and Boland. For
this purpose, we introduce the notion of a strong path. Two
non-adjacent vertices $u$ and $v$ are linked by a strong path if
either they have a common neighbor or they are the endpoints of
two vertex-disjoint chordless paths satisfying certain technical
conditions. (The complete definition is given in
Section~\ref{sec:strongpath}.) A \emph{strong asteroidal triple}
in a graph $G$ is an asteroidal triple such that each pair of
vertices of the triple is linked by a strong path in
$G$.

Our main result is the following theorem.
\begin{theorem}
  \label{th:directed}
  A chordal graph is a directed path graph if and only if it contains
  no strong asteroidal triple.
\end{theorem}
In Section \ref{sec:background}, we give the definitions and
background results needed to prove our theorems. In
Section~\ref{sec:interval}, we give a new proof of
Theorem~\ref{th:intervalgraph} based on clique trees. In
Section~\ref{sec:strongpath}, we define strong paths and establish
a property of strong paths in clique directed path trees (which are defined in Section \ref{sec:background}). In
Section~\ref{sec:directed}, we give a proof of
Theorem~\ref{th:directed} using the results of
Section~\ref{sec:strongpath}. In sections~\ref{sec:quad}
and~\ref{sec:forbidden}, we discuss asteroidal quadruples and
their relationship with graphs which are minimally not rooted path graph. Finally,
in Section~\ref{sec:conclusion}, we discuss new problems arising
from our work.

\section{Definitions and background}\label{sec:background}

In a graph $G$, a \emph{clique} is a set of pairwise adjacent
vertices.  Let $\mathcal Q(G)$ be the set of all (inclusionwise)
maximal cliques of $G$.  When there is no ambiguity we will write
$\mathcal Q$ instead of $\mathcal Q(G)$. If a vertex $u$ is adjacent
to a vertex $v$, we say that $u$ {\it sees} $v$; otherwise, we say $u$
{\it misses} $v$. A vertex in a graph $G$ is called \emph{universal}
if it sees every other vertex of $G$. Given a vertex $v$ and a set
$S$ of vertices, $v$ is called \emph{complete to $S$} if $v$ sees every
vertex of $S$.  Given two vertices $u$ and $v$ in a graph $G$, a
\emph{$\{u,v\}$-separator} is a set $S$ of vertices of $G$ such that
$u$ and $v$ lie in two different components of $G\setminus S$ and $S$
is minimal with this property.  A set is a \emph{separator} if it is a
$\{u,v\}$-separator for some $u$ and $v$ in $G$.  Let $\mathcal S(G)$ be the
set of separators of $G$.  When there is no ambiguity we will write
$\mathcal S$ instead of $\mathcal S(G)$.  A classical
result~\cite{HajSur58,Ber60} (see also~\cite{Gol04}) states that, in a
chordal graph $G$, every separator is a clique; moreover, if $S$ is a
separator, then there are at least two components of $G\setminus S$
that contain a vertex that is complete to $S$, and so $S$ is the
intersection of two maximal cliques.

A \emph{clique tree} $T$ of a graph $G$ is a tree whose vertices are
the members of $\mathcal Q$ and such that, for each vertex $v$ of $G$,
those members of $\mathcal Q$ that contain $v$ induce a subtree of
$T$, which we will denote by $T^v$.  A classical result~\cite{Gav74}
states that a graph is chordal if and only if it has a clique tree.  A
\emph{clique path tree} $T$ of $G$ is a clique tree of $G$ such that,
for each vertex $v$ of $G$, $T^v$ is a path.  Gavril \cite{Gav78} proved that a graph is a
path graph if and only if it has a clique path tree.  A \emph{clique
  directed path tree} $T$ of $G$ is a clique path tree of $G$ such
that edges of the tree $T$ are directed and for each vertex $v$ of
$G$, the subpath $T^v$ is a directed path. A \emph{clique rooted path
  tree} $T$ of $G$ is a clique directed path tree of $G$ such that
$T$ is a rooted tree.  Monma and Wei~\cite{MW86} proved that a graph is a
directed path graph if and only if it has a clique directed path
tree, and that a graph is a rooted path graph if and only if it
has a clique rooted path tree.  A \emph{clique path} $T$ of $G$ is
a clique tree of $G$ such that $T$ is a path. A graph is an
interval graph if and only if it has a clique path
\cite{FulGro65}.  These results allow us to consider only the
intersection models that are clique trees when studying the
properties of the graph classes.

For a clique tree $T$, the \emph{label} of an edge $QQ'$ of $T$ is
defined as $S_{QQ'}=Q\cap Q'$.  Note that every edge $QQ'$
satisfies $S_{QQ'} \in\mathcal S$; indeed, there exist vertices
$v\in Q\setminus Q'$ and $v'\in Q'\setminus Q$ such that the set
$S_{QQ'}$ is a $\{v, v'\}$-separator.  The number of times an
element $S$ of $\mathcal S$ appears as a label of an edge is equal
to $c-1$, where $c$ is the number of components of $G\setminus S$
that contain a vertex complete to $S$ \cite{Gav74,MacMac}.  Note
that this number is at least one and that it depends only on $S$
and not on $T$, so for a given $S\in \mathcal S$ it is the same in
every clique tree. For more information about clique trees and
chordal graphs, see \cite{Gol04, MacMac}.

Given $X \subseteq \mathcal Q$, let $G(X)$ denote the subgraph of
$G$ induced by all the vertices that appear in the members of $X$.
If $T$ is a clique tree of $G$, then $T[X]$ denotes the subtree of
$T$ of minimum size whose vertices contains $X$.  Note that if
$|X|=2$, then $T[X]$ is a path.  Given a subtree $T'$ of a clique
tree $T$ of $G$, let $\mathcal Q(T')$ be the set of vertices of
$T'$ and $\mathcal S(T')$ be the set of separators of $G(\mathcal
Q(T'))$. Note that $T'$ is a clique tree of $G(\mathcal Q(T'))$.

Given a set $z_1,\ldots,z_r$, $r\geq 2$, of pairwise non-adjacent
vertices of $G$, and a clique tree $T$ of $G$, the subtrees
$T^{z_i}$, $1\leq i\leq r$, are disjoint and we can define
$T(z_1,\ldots,z_r)$ the subtree of $T$ of minimum size that
contains at least one vertex of each $T^{z_i}$.  Clearly, the
number of leaves of $T(z_1,\ldots,z_r)$ is at most $r$. Moreover, if
$T(z_1,\ldots,z_r)$ has exactly $r$ leaves, then they can be
denoted by $Q_i$, $1\leq i\leq r$, with
$Q_{i}\cap\{z_1,\ldots,z_r\}=\{z_i\}$.

\section{Asteroidal triples}\label{sec:interval}

In this section, we give a proof of Theorem~\ref{th:intervalgraph}
using clique trees. First, we need the following lemma which is
folklore (for example, see \cite{LinMck98}.)

\begin{lemma}
  \label{lem:asteroidaltriple}
  Let $G$ be a chordal graph and $z_1,z_2,z_3$ three vertices that form an
  asteroidal triple, then for every clique tree $T$ of $G$, the
  subtree $T(z_1,z_2,z_3)$ has exactly $3$ leaves.
\end{lemma}

\begin{proof}
  Suppose the subtree $T(z_1,z_2,z_3)$ is a path. For $1\leq i\leq
  3$, let $Q_i$ be a vertex of $T(z_1,z_2,z_3)$ containing $z_i$ (they
  are all distinct as they are cliques of $G$ and $z_1,z_2,z_3$ are
  not adjacent). We can assume that $Q_1,Q_2,Q_3$ appear in this
  order along the path $T(z_1,z_2,z_3)$. Vertices $z_1$ and $z_3$ are in
  two different components of the graph $G\backslash Q_2$ so every
  path that goes from $z_1$ to $z_3$ has to use a vertex of $Q_2$ and
  thus a neighbor of $z_2$, contradicting the fact that $z_1,z_2,z_3$ is
  an asteroidal triple.
\end{proof}

A consequence of Lemma~\ref{lem:asteroidaltriple} is that an
interval graph does not contain an asteroidal triple. Lekkerkerker
and Boland~\cite{LB62} proved that the converse is also true.
Halin~\cite{Hal82} gave a short proof of
Theorem~\ref{th:intervalgraph}. Unfortunately, this proof is hard
to follow as it uses the so called \emph{prime graph
decomposition}.  In the book~\cite{MacMac}, McKee and McMorris
shorten the proof of Halin by considering the clique tree, but
this proof is incomplete. 

It seems to be easier to understand Halin's proof in terms of clique trees
rather than in terms of his prime graph decomposition. We are
going to give such a proof.  The ideas of this proof have been
generalized and extensively used in~\cite{LMP08} to obtain a
forbidden subgraph characterization of path graphs.

\

\noindent \emph{Proof of Theorem~\ref{th:intervalgraph}.}\
\emph{($\Longrightarrow$)} Suppose $G$ is an interval graph and
$z_1,z_2,z_3$ is an asteroidal triple of $G$.  Let $T$ be a clique
path of $G$. By Lemma~\ref{lem:asteroidaltriple}, the subtree
$T(z_1,z_2,z_3)$ has exactly $3$ leaves, so $T$ is not a path, a
contradiction.

\emph{($\Longleftarrow$)} Suppose that $G$ is a chordal graph
containing no asteroidal triple and that $G$ is a minimal non-interval graph. Let $T$ be any clique tree of $G$. The graph $G$
is not an interval graph, so $T$ has at least three distinct and
non-adjacent leaves $Q_1,Q_2,Q_3$. For $i=1,2,3$, let $Q_i'$ be
the neighbor of $Q_i$ on $T$. (The vertices $Q_i'$ are distinct
from $Q_1,Q_2,Q_3$ but not necessarily pairwise distinct.)  For
$i=1,2,3$, let $z_i\in Q_i\backslash Q_i'$ and $S_i=Q_i\cap Q_i'$.
Vertices $z_1,z_2,z_3$ are three non-adjacent vertices that do not
form an asteroidal triple. So, by symmetry, we can assume that
every path that goes from $z_2$ to $z_3$ use some vertices in
$N(z_1)$.

We claim that there is an edge of $T[Q_2,Q_3]$ such that $z_1$ is
complete to the label of the edge. (Recall that a label is a set of vertices. So here,
$z_1$ is complete to the set of vertices which are the label of the edge.)
Suppose, on the contrary, that there is no
edge of $T[Q_2, Q_3]$ such that $z_1$ is complete to its label.
Then, in each label of edges of $T[Q_2, Q_3]$, one can select a
vertex of $G$ that is not adjacent to $z_1$.  The set of
selected vertices forms a path from $z_2$ to $z_3$ that uses no
vertex from $N(z_1)$, a contradiction. So $z_1$ is complete to the
label $S_0$ of an edge of $T[Q_2,Q_3]$, so $S_0\subseteq S_1$. The
vertex $Q_1$ is not on $T[Q_2,Q_3]$, so $S_0$ and $S_1$ are labels
of two different edges of $T$, even if they may be equal.

Let $P$ be a clique path of $G(\mathcal Q(T)\backslash \{Q_1\})$.  Let
$T'$ be the clique tree of $G$ obtained from $P$ by adding vertex
$Q_1$ and edge $Q_1Q_1'$.  For the two clique trees $T$ and $T'$ of
$G$, the number of times a label appears in each clique tree is
the same, so $S_0,S_1$ are labels of two different edges of $T'$ and
so $S_0$ is the label of an edge of $P$.

Let $P'$ be the maximal subpath of $P$ that contains $Q_1'$ and
such that no label of edges of $P'$ is a subset of $S_1$.  Let
$T_0$ be the clique tree of $G(\mathcal Q(P')\cup \{Q_1\})$
obtained from $P'$ by adding vertex $Q_1$ and edge $Q_1Q_1'$. As
$Q_1$ is a leaf of $T_0$, every label of $T_0$ that is a subset of
$Q_1$ is a subset of $S_1$. Since only one label of $T_0$ is a
subset of $S_1$, only one label of $T_0$ is a subset of $Q_1$. The
label $S_0$ is a subset of $S_1$, and so $P'$ has strictly fewer
vertices than $P$. So $G(\mathcal Q(P')\cup \{Q_1\})$ is an
interval graph. Let $P_0$ be a clique path of this graph.  For the
two clique trees $T_0$, $P_0$ of $G(\mathcal Q(P')\cup \{Q_1\})$,
the number of times a label appears in each clique tree is
the same, so only one label of $P_0$ is a subset of $Q_1$. So
$Q_1$ is a leaf of $P_0$.

The path $P'$ is a proper subpath of $P$, so $P\backslash P'$ is
either a path or the union of two paths.

\emph{Case 1 : $P\backslash P'$ is a path.}  Let $P_1$ be the path
$P\backslash P'$. Let $L$ be the leaf of $P_1$ such that there
exists a vertex $L'$ in $P'$ with $LL'$ being an edge of $P$. By
definition of $P'$, the label $S_{LL'}$ of $LL'$ is included in
$S_1$. So $P_0$ and $P_1$ can be linked by the edge $LQ_1$ to
obtain a clique path of $G$, a contradiction.

\emph{Case 2 : $P\backslash P'$ is the union of two paths.}  Let $P_1,
P_2$ the paths of $P\backslash P'$. Let $L_i$ be the leaf of $P_i$
such that there exists a vertex $L_i'$ in $P'$ with $L_iL'_{i}$ being
an edge of $P$. By definition of $P'$, labels $S_{L_iL_i'}$ of
$L_iL_i'$ are subsets of $S_1$. Let $L_0$ be the leaf of $P_0$ that is
different from $Q_1$. The vertex $L_0$ is a vertex of $P'$, so it is
either on $P'[Q_1',L_1']$ or on $P'[Q_1',L_2']$. Suppose, by symmetry,
that $L_0$ is on $P'[Q_1',L_2']$. Since every vertex of the path
$P'[Q_1',L_2']$ contains $S_{L_2L_2'}$, vertex $L_0$ contains
$S_{L_2L_2'}$. So $P_0$, $P_1$ and $P_2$ can be linked by edges
$L_1Q_1$, $L_2L_0$ to obtain a clique path of $G$, a contradiction.
{\hfil$\Box$\vspace{1em}

\section{Strong paths and clique directed path
trees}\label{sec:strongpath}

Two non-adjacent vertices $u$ and $v$ are linked by a \emph{strong
path} if either they have a common neighbor, or there exist three
sets of distinct vertices
$X=\{x_1,\ldots,x_r\},Y=\{y_1,\ldots,y_s\},Z$, ($r,s\geq 2$,
$|Z|\geq 0$), such that $u$-$x_1$-$\cdots$-$x_r$-$v$ and
$u$-$y_1$-$\cdots$-$y_s$-$v$ are two chordless paths 
where if
$x_i,x_{i+1},y_j,y_{j+1}$ ($1\leq i<r$, $1\leq j<s$) is a clique of
size four, one of the following is satisfied with
$\{l_1,l_2\}=\{x_i,y_j\}$ and $\{r_1,r_2\}=\{x_{i+1},y_{j+1}\}$:

\begin{itemize}
\item\emph{Attachment of type 1 on $\{l_1,l_2\},\{r_1,r_2\}$ :}
There
  exist two non-adjacent vertices $z,z'$ of $Z$ such that vertex $z$
  sees $l_1,l_2,r_1$ and not $r_2$ and vertex $z'$ sees $r_1,r_2,l_1$ and
  not $l_2$.
\item\emph{Attachment of type 2 on $\{l_1,l_2\},\{r_1,r_2\}$ :}
There
  exist $4t+3$ ($t\geq0$) vertices
  $z_1,\ldots,z_{2t+1},z'_1,\ldots,$ $z'_{2t+2}$ of $Z$ such that
  vertices $l_1,l_2,r_1,r_2,z_1,\ldots,z_{2k+1}$ form a clique $Q$,
  vertices $z_1,\ldots,z_{2t+1}$ sees exactly $l_1,l_2,r_1,r_2$ on
  $X\cup Y\cup\{u,v\}$, vertices $z'_1,\ldots,z'_{2t+2}$ form a stable
  set, and vertex $z'_k$ ($1\leq k\leq 2t+2$) sees exactly $z_{k-1},z_k$
  on $Q\cup X\cup Y\cup\{u,v\}$ (with $z_0=l_1$ and $z_{2t+2}=r_1$).
\end{itemize}

The graphs of Figure~\ref{fig:strongpath} are examples of chordal
graphs in which vertices $u$ and $v$ are linked by a strong path.

\begin{figure}[h!]
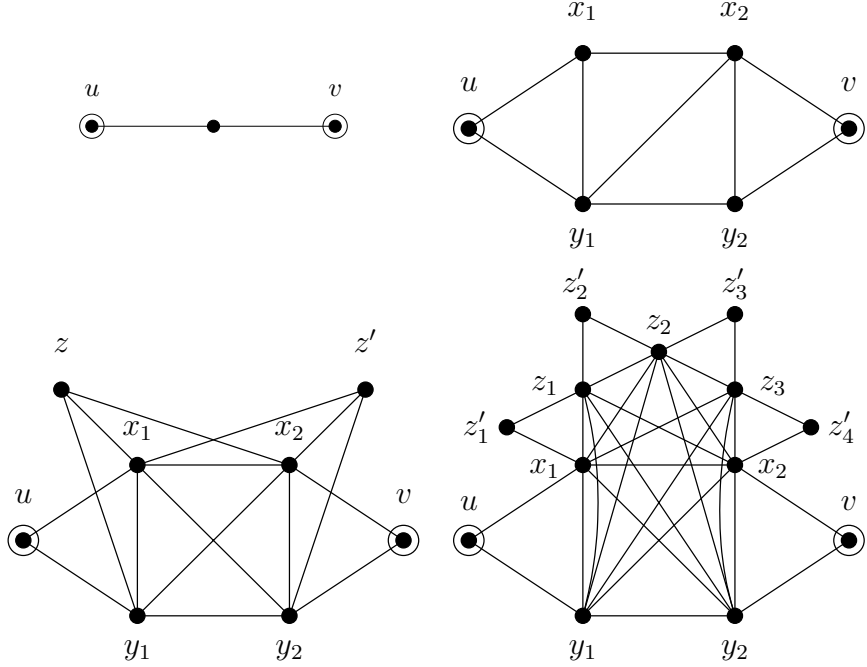

  \centering
  \begin{tabular}{cc}
\input{zfig-strongpath-1.pstex_t} &
\input{zfig-strongpath-2.pstex_t} \\
\input{zfig-strongpath-3.pstex_t} &
\input{zfig-strongpath-4.pstex_t}
  \end{tabular}
  \caption{Examples of strong paths}
  \label{fig:strongpath}
\end{figure}

Strong paths are interesting when considering directed path graphs
because of the following lemma.

\begin{lemma}
\label{lem:strongpathdirection}
Let $G$ be a directed path graph and $u$ and $v$ two non-adjacent vertices
that are linked by a strong path, then for every clique directed path
tree $T$ of $G$, the subpath $T(u,v)$ is a directed path.
\end{lemma}

\begin{proof}
  Suppose on the contrary that $T$ is a clique directed tree of $G$
  such that $T(u,v)$ is not a directed path. Let $Q_u$ and $Q_v$ be
  the two extremities of $T(u,v)$ with $u\in Q_u$ and $v\in Q_v$.  By
  assumption, the subpath $T(u,v)$ contains two edges not oriented in
  the same direction. Let $Q_1,Q_2,Q_3$ be three consecutive vertices
  of $T(u,v)$ such that $Q_u,Q_1,Q_2,Q_3,Q_v$ appears in this order
  along $T(u,v)$ (we may have $Q_u=Q_1$, $Q_3=Q_v$) and edges $Q_1Q_2$
  and $Q_2Q_3$ are not oriented in the same direction. By reversing
  all the edges of $T$ if necessary, we may assume that $Q_1\rightarrow Q_2$
  and $Q_2\leftarrow Q_3$ (where $Q_1\rightarrow Q_2$ mean there is a edge directed from $Q_1$
  to $Q_2$). As $T$ is a clique
  directed path tree, $S_{Q_1Q_2}\cap S_{Q_2Q_3}=\emptyset$.

  Suppose that there exists $w\in N(u)\cap N(v)$. Then $w$ is not in
  $S_{Q_1Q_2}$ or not in $S_{Q_2Q_3}$. By symmetry, we may assume that
  $w\notin S_{Q_1Q_2}$. Then $u$ and $v$ are in two components of
  $G\backslash S_{Q_1Q_2}$, a contradiction because $u$-$w$-$v$ is a
  path. So, we have $N(u)\cap N(v)= \emptyset$ and there exists a
  strong path with $X=\{x_1,\ldots,x_r\},Y=\{y_1,\ldots,y_s\},Z$,
  ($r,s\geq 2$, $|Z|\geq 0$).

  We claim that every label of the edges of $T(u,v)$ contains at least
  one vertex from $X$ and at least one vertex from $Y$. Suppose on the
  contrary that there exists an edge $QQ'$ of $T(u,v)$ such that
  $S_{QQ'}\cap X=\emptyset$. Then $u$ and $v$ are in two different
  components of $G\backslash S_{QQ'}$, contradicting the fact that
  $u$-$X$-$v$ is a path. The case for $Y$ is similar and thus our claim
  holds.

  Let $i$ and $j$ be the maximum subscripts with $1\leq i\leq r$, $1\leq j
  \leq s$ such that $x_i,y_j \in S_{Q_1Q_2}$.  As $S_{Q_1Q_2}\cap
  S_{Q_2Q_3}=\emptyset$, we have $x_i,y_j \in S_{Q_1Q_2}\backslash
  S_{Q_2Q_3}$. Vertex $v$ is not in $Q_2$ by definition of $T(u,v)$,
  so not in $S_{Q_2Q_3}$. In $G\backslash S_{Q_2Q_3}$, vertices $x_i,
  y_j$ are not in the same component as $v$, so $x_i, y_j$ are not
  adjacent to $v$ and $i<r$, $j<s$.

  The set $S_{Q_2Q_3}$ contains at least one vertex from $X$ and at
  least one vertex from $Y$. Vertices of $S_{Q_2Q_3}$ are in $Q_2$, and thus are
  adjacent to both $x_i,y_j$. As $X$ and $Y$ are chordless paths,
  $S_{Q_2Q_3}$ contains at least one of $x_{i-1},x_{i+1}$ and at least
  one of $y_{j-1},y_{j+1}$.

  Suppose that $S_{Q_2Q_3}$ contains $x_{i-1}$ (so $i>2$). Then
  $x_{i-1}$ is not in $S_{Q_1Q_2}$. Because $S_{Q_1Q_2}$ contains
  $x_i$ and $u$-$x_1$-$\cdots$-$x_{i-1}$-$x_{i}$ is a chordless path,
  $S_{Q_1Q_2}$ contains no vertex of $\{u,x_1,\ldots,x_{i-2}\}$.  Then
  $u$ and $x_{i-1}$ are in two different components of $G\backslash
  S_{Q_1Q_2}$, contradicting the fact that $u$-$x_1$-$\cdots$-$x_{i-1}$ is a
  path.  So, we have $x_{i-1}\notin S_{Q_2Q_3}$, and $x_{i+1}\in
  S_{Q_2Q_3}$.  Similarly, we have $y_{j+1}\in S_{Q_2Q_3}$.

  Now, the vertices $x_i,y_j,x_{i+1},y_{j+1}$ form a clique of size four in
  $X\cup Y$. So, there are vertices of $Z$ forming an attachment of
  type $1$ or $2$ on $\{x_i,y_j\},\{x_{i+1},y_{j+1}\}$.

  Suppose first that there is an attachment of type 1.  Then, by
  symmetry, we may assume that there exist two non-adjacent vertices
  $z,z'$ of $Z$ such that $z$ sees $x_i,y_j,x_{i+1}$ and not $y_{j+1}$
  and $z'$ sees $x_i,x_{i+1},y_{j+1}$ and not $y_{j}$.  Let $Q_z$ be a
  vertex of $T$ containing $z,x_i,y_j,x_{i+1}$ and $Q_{z'}$ a vertex
  of $T$ containing $z',x_i,x_{i+1},y_{j+1}$. All of
  $Q_1,Q_2,Q_{z},Q_{z'}$ contain $x_i$ so there are all vertices of
  the path $T^{x_i}$. Vertex $Q_1$ is not between $Q_z$ and $Q_2$ or
  between $Q_{z'}$ and $Q_2$ along this path, since otherwise $S_{Q_1Q_2}$ contains
  $x_{i+1}$. Vertex $Q_z$ is not between $Q_z'$ and $Q_2$, since otherwise $z$
  sees $y_{j+1}$. So vertices $Q_1,Q_2,Q_{z'},Q_z$ appear in this
  order along the path $T^{x_i}$, but then $z'$ sees $y_j$, a
  contradiction.

  Suppose now that there is an attachment of type 2. Then, by symmetry,
  we can assume that there exist $4t+3$ ($t\geq0$) vertices
  $z_1,\ldots,z_{2t+1},z'_1,\ldots,z'_{2t+2}$ of $Z$ such that
  vertices $x_i,y_j,x_{i+1},y_{j+1},z_1,\ldots,$ $z_{2t+1}$ form a clique
  $Q$, vertices $z_1,\ldots,z_{2t+1}$ see exactly
  $x_i,y_j,x_{i+1},y_{j+1}$ on $X\cup Y\cup\{u,v\}$, vertices
  $z'_1,\ldots,z'_{2t+2}$ form a stable set, vertex $z'_k$ ($1\leq
  k\leq 2t+2$) sees exactly $z_{k-1}$ and $z_k$ on $Q\cup X\cup
  Y\cup\{u,v\}$ (with $z_0=x_i$ and $z_{2t+2}=x_{i+1}$).  Let $K$ be a
  vertex of $T$ containing $Q$ (we may have $K=Q_2$) and for $1\leq k\leq
  2t+2$, let $Z_k$ be a vertex of $T$ containing
  $z'_k,z_{k-1},z_k$.

  Vertices $Q_1,Q_2,K$ are all on the path $T^{x_i}$. If $Q_1$ is
  between $K$ and $Q_2$, then $S_{Q_1Q_2}$ contains $y_{j+1}$, a
  contradiction. So $Q_1,Q_2,K$ appear in this order along
  $T^{x_i}$. Vertex $Z_1$ is also on $T^{x_i}$.  Vertex $Z_1$ is not between
  $Q_1$ and $K$ as it does not contain $y_j$.

  Suppose that $Z_1,Q_1,Q_2,K$ appear in this order along $T^{x_i}$.
  If $Z_1$ is on $T^{y_j}$, then $z'_1$ sees $y_j$, a
  contradiction. So $Z_1$ is not on $T^{y_j}$. Let $Q'_1$ be the
  clique of $T(Z_1,Q_1)$ nearest to $Z_1$ that contains $y_j$
  (we may have $Q'_1=Q_1$). If $Q'_1=Q_u$, then $z_1$ sees $u$, a
  contradiction. So $Q'_1\neq Q_u$.  Let $Z'_1$ be the neighbor of
  $Q'_1$ on $T(Z_1,Q'_1)$. If $Z'_1$ is a vertex of $T(u,v)$, then
  $S_{Z'_1Q'_1}$ contains a vertex of $Y$ that sees $y_j$ and that is
  different from $y_{j+1}$, so $S_{Z'_1Q'_1}$ contains $y_{j-1}$ and
  $z_1$ sees $y_{j-1}$, a contradiction. So $Z'_1$ is not a vertex of
  $T(u,v)$.  Let $Q''_1$ be the clique of $T(Q'_1,Q_1)$ nearest
  to $Q'_1$ that is in $T(u,v)$ (we may have $Q''_1=Q_1$ or
  $Q''_1=Q'_1$ or both). If $Q''_1=Q_u$, then $z_1$ sees $u$, a
  contradiction. Let $Q'_u$ be the neighbor of $Q''_1$ on $T(u,v)$ that
  is not on $T(Z_1,Q_2)$.  Then $S_{Q'_uQ''_1}$ contains a vertex of
  $X$ that sees $x_i$ and that is different from $x_{i+1}$, so
  $S_{Q'_uQ''_1}$ contains $x_{i-1}$ and $z_1$ sees $x_{i-1}$, a
  contradiction.  So $Q_1,Q_2,K,Z_1$ appear in this order along
  $T^{x_i}$. Similarly, $Q_3,Q_2,K,Z_{2t+2}$ appear in this order along
  $T^{x_{i+1}}$.

  For every $k$ and $\ell$, $1\leq k < \ell\leq 2t+2$, vertices $Z_k,K,Z_\ell$
  appear in this order along $T[K,Z_k,Z_\ell]$, since otherwise, $Z_k$
  contains $z_\ell$ or $Z_\ell$ contains $z_{k-1}$.  Vertices $Q_1,K,Z_1$
  appear in this order along $T^{x_i}$ and $Q_1\rightarrow Q_2$, so
  $T[K,Z_1]$ is directed from $K$ to $Z_1$.  Vertices $Z_1,K,Z_2$
  appear in this order along $T^{z_1}$, so $T[K,Z_2]$ is directed
  from $Z_2$ to $K$. And so on, for $2\leq k \leq 2t+2$, vertices
  $Z_{k-1},K,Z_k$ appear in this order along $T^{z_k}$, so $T[K,Z_k]$
  is directed from $K$ to $Z_k$ when $k$ is odd and from $Z_k$ to $K$
  when $k$ is even. So $T[K,Z_{2t+2}]$ is directed from $Z_{2t+2}$ to
  $K$. Vertices $Q_3,Q_2,K,Z_{2t+2}$ appear in this order along
  $T^{x_{i+1}}$, so $T[K,Q_3]$ is directed from $K$ to $Q_3$,
  contradicting $Q_2\leftarrow Q_3$.
\end{proof}

\section{Asteroidal triples in directed path
graphs}\label{sec:directed}

The graph of Figure~\ref{fig:asteroidaltriple} is a directed path
graph that is minimally not an interval graph; in fact, it is a rooted
path graph. (It is the graph $F_{18}$ of Figure~\ref{fig:interval}.)
So, directed path graphs may contain asteroidal triples. But one can
define a particular type of asteroidal triple that is forbidden in
directed path graphs.  Recall from Section~\ref{sec:intro} that a
\emph{strong asteroidal triple} in a graph $G$ is an asteroidal triple
such that each pair of vertices of the triple are linked by a strong
path in $G$.  The graph of Figure~\ref{fig:strongasteroidaltriple1} is
example of a graph that minimally contains a strong asteroidal
triple. This graph is a path graph which is minimally not a directed
path graph. (It is the graph $F_{17}(6)$ of Figure~\ref{fig:directed},
and also $F_{21}(6)$ of Figure~\ref{fig:interval}.)

\begin{figure}[h]
  \centering
\input{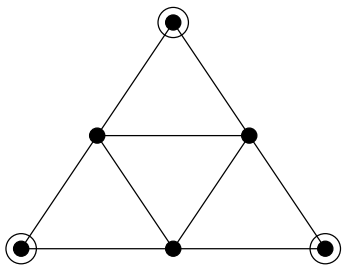}
  \caption{Graph containing a strong asteroidal triple}
  \label{fig:strongasteroidaltriple1}
\end{figure}

The graph of Figure~\ref{fig:strongasteroidaltriple2} is another
example of a graph that minimally contains a strong asteroidal
triple. This graph is interesting as it shows that sometimes the
path between two vertices of the asteroidal triple that avoids the
neighborhood of the third must contain some vertices outside the
strong path. The only strong path linking $2$ and $3$ is
$X=\{x_1,x_2\},Y=\{y_1,y_2\},Z=\emptyset$ and the only path
between $2$ and $3$ that avoids the neighborhood of $1$ is
$y_1$-$t$-$x_2$. This graph is a chordal graph which is minimally not a path
graph. (It is the graph $F_{10}(8)$ of Figures~\ref{fig:path},
\ref{fig:directed} and $F_{21}(8)$ of Figure~\ref{fig:interval}.)

\begin{figure}[h]
  \centering
\input{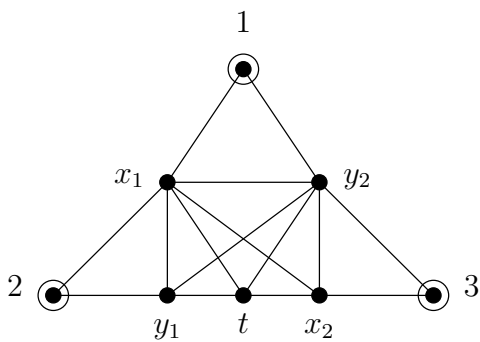}
  \caption{Graph containing a strong asteroidal triple}
  \label{fig:strongasteroidaltriple2}
\end{figure}

The graph of Figure~\ref{fig:asteroidaltriple} is an example of a
graph that contains an asteroidal triple that is not strong as for
two vertices of the asteroidal triple, there is no common neighbor
and no pair of disjoint paths between them.  The graph of
Figure~\ref{fig:asteroidaltriplenotstrong} is another example of a
graph that contains an asteroidal triple that is not strong. In
this graph, there exist two disjoint paths
$\{x_1,x_2\},\{y_1,y_2\}$ between $2$ and $3$ but
$x_1,x_2,y_1,y_2$ is a clique of size four and there are no
vertices that can play the role of $Z$ in the definition of strong
path. This graph is a rooted path graph which is minimally not an interval
graph. (It is the graph $F_{21}(7)$ of Figure~\ref{fig:interval}.)

\begin{figure}[h]
  \centering
\input{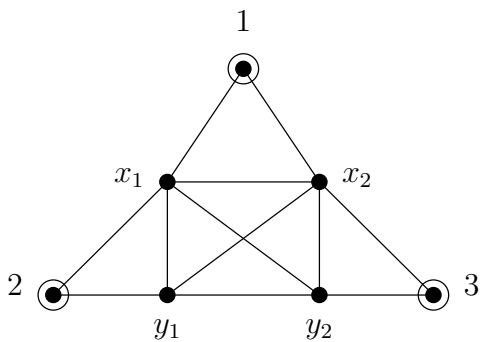}
  \caption{Graph containing an asteroidal triple that is not strong}
  \label{fig:asteroidaltriplenotstrong}
\end{figure}

We will now prove Theorem~\ref{th:directed} which gives a
characterization of directed path graphs using strong asteroidal
triples.

 {\it Proof of Theorem \ref{th:directed}.}
  \emph{($\Longrightarrow$)} Suppose that $G$ is a directed path graph
  and $z_1,z_2,z_3$ is a strong asteroidal triple of $G$. Let $T$ be a
  clique directed path tree of $G$. By
  Lemma~\ref{lem:asteroidaltriple}, $T(z_1,z_2,z_3)$ has exactly $3$
  leaves $Q_i$, $1\leq i\leq 3$, with $z_i\in Q_i$.  By
  Lemma~\ref{lem:strongpathdirection}, $T(z_1,z_2)$, $T(z_2,z_3)$,
  $T(z_3,z_1)$ are directed paths of $T(z_1,z_2,z_3)$. Suppose, by
  symmetry, that $T(z_1,z_2)$ is directed from $Q_1$ to $Q_2$. Then
  $T(z_1,z_3)$ is directed from $Q_1$ to $Q_3$, but then $T(z_2,z_3)$
  is not a directed path, a contradiction.

  \emph{($\Longleftarrow$)} All chordal graphs of
  Figure~\ref{fig:directed} contain a strong asteroidal triple.
  The graphs $F_1$, $F_3$, $F_4$, $F_5(n)_{n\geq 7}$ are obtained from a
  graph containing an asteroidal triple by adding a universal vertex;
  this universal vertex forms a strong path linking each pair of
  vertices of the asteroidal triple.  In the graphs $F_6$, $F_7$, $F_9$,
  $F_{10}(n)_{n\geq 8}$, the strong paths are either a common neighbor or
  four vertices without attachment.  In the graphs $F_{13}(4k+1)_{k\geq
    2}$, $F_{15}(4k+2)_{k\geq 2}$, $F_{16}(4k+3)_{k\geq 2}$,
  $F_{17}(4k+2)_{k\geq 1}$, the  strong paths are either a common neighbor,
  four vertices without attachment, or four vertices plus attachment of
  type $1$ or $2$.  So if $G$ is a chordal graph containing no strong
  asteroidal triple, it contains no $F_1$, $F_3$, $F_4$,
  $F_5(n)_{n\geq 7}$, $F_6$, $F_7$, $F_9$, $F_{10}(n)_{n\geq 8}$,
  $F_{13}(4k+1)_{k\geq 2}$, $F_{15}(4k+2)_{k\geq 2}$,
  $F_{16}(4k+3)_{k\geq 2}$, $F_{17}(4k+2)_{k\geq 1}$, and so it is a
  directed path graph by the result of Panda~\cite{Pan99}.
\hfill $\Box$

In the proof of Theorem~\ref{th:directed}, we use the list of
forbidden subgraph obtained by Panda~\cite{Pan99}. It would be
nice to find a simple and direct proof of this result, similar to the
proof of Theorem~\ref{th:intervalgraph} presented in
Section~\ref{sec:interval}.

A corollary of Theorem~\ref{th:directed} and~\cite{Pan99} is the
following.

\begin{corollary}
\label{cor:strongasteroidaltriple}
  The chordal graphs that minimally contain a strong asteroidal triple
  are the graphs $F_1$, $F_3$, $F_4$, $F_5(n)_{n\geq 7}$, $F_6$, $F_7$,
  $F_9$, $F_{10}(n)_{n\geq 8}$, $F_{13}(4k+1)_{k\geq 2}$,
  $F_{15}(4k+2)_{k\geq 2}$, $F_{16}(4k+3)_{k\geq 2}$,
  $F_{17}(4k+2)_{k\geq 1}$.
\end{corollary}

One can notice that only \emph{short strong paths} are used in the
proof of Theorem~\ref{th:directed} (where \emph{short} means
either a common neighbor or $|X|=|Y|=2$). So if one is interested
only in characterizing directed path graphs, there is no need to
define \emph{long strong path} (where \emph{long} means one of
$|X|$, $|Y|$ is at least $3$). As we will see in next sections,
long strong paths are useful for rooted path graphs.

\section{Asteroidal quadruples in rooted path
graphs}\label{sec:quad}

The notion of asteroidal triple can be generalized to four
vertices. An \emph{asteroidal quadruple} in a graph $G$ is a set
of four vertices such that any three of them is an asteroidal
triple.  The graph of Figure~\ref{fig:asteroidalquadruple} is an
example of a graph that minimally contains an asteroidal
quadruple.

\begin{figure}[h]
  \centering
\includegraphics[scale=1]{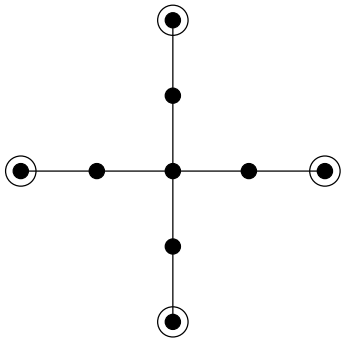}
  \caption{Graph containing an asteroidal quadruple}
  \label{fig:asteroidalquadruple}
\end{figure}

The following lemma is analogous to
Lemma~\ref{lem:asteroidaltriple} for asteroidal quadruple (see
also \cite{LinMck98}).

\begin{lemma}
  \label{lem:asteroidalquadruple}
  Let $G$ be a chordal graph and let $z_1,z_2,z_3,z_4$ be four vertices that form an
  asteroidal quadruple. Then for every clique tree $T$ of $G$, the
  subtree $T(z_1,z_2,z_3,z_4)$ has exactly 4 leaves.
\end{lemma}

\begin{proof}
  Suppose that the subtree $T(z_1,z_2,z_3,z_4)$ has fewer than
  $4$ leaves. Then there is at least one $z_i$ that is not in a leaf.
  Suppose by symmetry that $z_4$ is not in a leave. Let $Q_4$ be a vertex
  of $T(z_1,z_2,z_3,z_4)$ that contains $z_4$.  Then
  $T(z_1,z_2,z_3,z_4)=T(z_1,z_2,z_3)$ and by
  Lemma~\ref{lem:asteroidaltriple}, $T(z_1,z_2,z_3)$ has exactly $3$
  leaves. Vertex $Q_4$ is either on $T(z_1,z_2)$ or $T(z_1,z_3)$.
  Suppose by symmetry that $Q_4$ is a vertex of $T(z_1,z_2)$. Then
  $T(z_1,z_2,z_4)=T(z_1,z_2)$ is a path, contradicting
  Lemma~\ref{lem:asteroidaltriple}.
\end{proof}

The graph of Figure~\ref{fig:asteroidalquadruple} is a rooted path
graph, so a rooted path graph may contain asteroidal quadruples.
But one can define a particular type of asteroidal quadruple that
is forbidden in rooted path graphs.

One can try to use the notion of strong asteroidal triple to
define a \emph{strong asteroidal quadruple} as a set of four
vertices such that any three of them is an strong asteroidal
triple. This is not interesting for our purpose because then every
graph that contains a strong asteroidal quadruple also contains a
strong asteroidal triple. And, by Theorem~\ref{th:directed}, we
already know that directed path graphs and thus rooted path graphs
contain no strong asteroidal triple.

One can define another four-vertex variant of asteroidal triple
that will be useful.  A \emph{weak asteroidal triple} in a
graph $G$ is an asteroidal triple such that two vertices
of the asteroidal triple are linked by a strong path in $G$.
The difference from the definition of a strong asteroidal
triple is that we do not expect that there is a strong path
linking any two of the three vertices but just linking two of
them.

Now we can generalize this notion to four vertices.  A \emph{weak
asteroidal quadruple} is a set of four vertices such that any
three of them is a weak asteroidal triple. Weak asteroidal
quadruples are interesting when considering directed path graphs
because of the following theorem.

\begin{theorem}
  \label{th:rootedpathgraph}
  A rooted path graph contains no weak asteroidal quadruple.
\end{theorem}

\begin{proof}
  Suppose that $G$ is a rooted path graph and $z_1,z_2,z_3,z_4$ is a
  weak asteroidal quadruple of $G$. Let $T$ be a clique rooted path
  tree of $G$. By Lemma~\ref{lem:asteroidalquadruple},
  $T(z_1,z_2,z_3,z_4)$ has exactly $4$ leaves $Q_i$, $1\leq i\leq 4$,
  with $z_i\in Q_i$. A subtree of a rooted tree is a rooted tree,
  so
  $T(z_1,z_2,z_3,z_4)$ is also rooted.

  By definition of weak asteroidal quadruple, the three vertices
  $z_1,z_2,z_3$ form a weak asteroidal triple, so two of them are
  linked by a strong path.  Suppose, by symmetry, that there is a
  strong path linking $z_1$ and $z_2$. Then, by
  Lemma~\ref{lem:strongpathdirection}, $T(z_1,z_2)$ is a directed
  path. So, $Q_1$ or $Q_2$ is the root of
  $T(z_1,z_2,z_3,z_4)$. Suppose by symmetry that $Q_1$ is the root of
  $T(z_1,z_2,z_3,z_4)$.

  The three vertices $z_2,z_3,z_4$ also form a weak asteroidal triple,
  so there is a strong path linking two of them. Let $z_i$ and $z_j$ be two
  vertices of $z_2,z_3,z_4$ that are linked by a strong path.  Then,
  by Lemma~\ref{lem:strongpathdirection}, $T(z_i,z_j)$ is a directed
  path; so $Q_i$ or $Q_j$ is the root of $T(z_1,z_2,z_3,z_4)$, a
  contradiction.
\end{proof}

By Theorems~\ref{th:directed} and~\ref{th:rootedpathgraph}, we
know that a rooted path graph contains no hole, no strong
asteroidal triple and no weak asteroidal quadruple. We conjecture
that the converse is also true.

\begin{conjecture}
 \label{conj:rootedpathgraph}
  A chordal graph is a rooted path graph if and only if it contains no
  strong asteroidal triple and no weak asteroidal quadruple.
\end{conjecture}

If Conjecture~\ref{conj:rootedpathgraph} is true, it will give a
characterization of rooted path graphs analogous to our
Theorem~\ref{th:directed} on directed path graphs and to Lekkerkerker and Boland's characterization of interval graphs.

\section{Forbidden subgraphs of rooted path graphs?}\label{sec:forbidden}

Recall that the problem of finding the forbidden induced subgraph
characterization is solved for interval graphs, path graphs, and
directed path graphs, but not for rooted path graphs.  Whether
Conjecture~\ref{conj:rootedpathgraph} is true or false,
Theorems~\ref{th:directed} and~\ref{th:rootedpathgraph} can be
used to obtain many graphs that are minimally not rooted path
graphs. What are the graphs that contain a strong asteroidal triple
or a weak asteroidal quadruple and that are minimally not rooted
path graphs?

The graphs of Figure~\ref{fig:weakasteroidalquadruple} are
examples of graphs that minimally contain a weak asteroidal
quadruple. They are also minimally not rooted path graphs. They
are of particular interest since together with $F_{17}(6)$ (see
Figure~\ref{fig:strongasteroidaltriple1}) they show that rooted
path graphs contain no suns. A \emph{sun} is the graph obtained by
taking a clique on vertices $a_1, a_2, \ldots, a_k$ for some $k
\geq 3$, a stable set on vertices $s_1, s_2, \ldots, s_k$, and
adding edges $s_i a_i, s_i a_{i+1}$ for all $i$, with the
subscripts taken modulo $k$. Farber showed \cite{Far83} that a
graph is strongly chordal if and only if it is chordal and does
not contain a sun. Thus, rooted path graphs are a subclass of
strongly chordal graphs.

\begin{figure}[h]
  \centering
\begin{tabular}{ccc}
\includegraphics[scale=1]{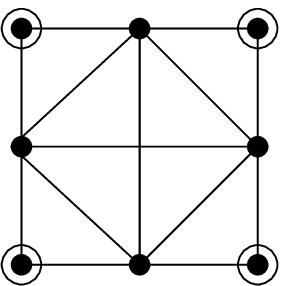} \ &\
\includegraphics[scale=1]{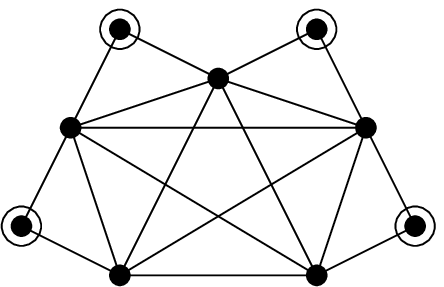} \ &\
\includegraphics[scale=1]{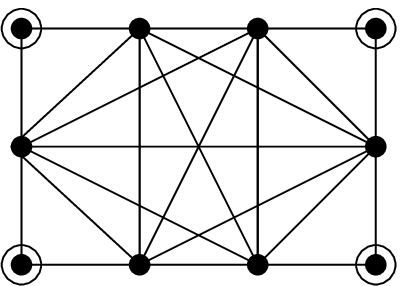} \\
$F_{22}$ & $F_{23}$ & $F_{24}$ \\
\end{tabular}
\caption{Examples of graphs which are minimally not rooted path graphs and contain a weak
  asteroidal quadruple}
  \label{fig:weakasteroidalquadruple}
\end{figure}

Corollary~\ref{cor:strongasteroidaltriple} gives the list of
chordal graphs that minimally contain a strong asteroidal triple.
One may notice that $F_{13}(4k+1)_{k\geq 3}$, $F_{15}(4k+2)_{k\geq
2}$, $F_{16}(4k+3)_{k\geq 2}$, and $F_{17}(4k+2)_{k\geq 2}$ all
strictly contain a weak asteroidal quadruple. All the other
chordal graphs that minimally contain a strong asteroidal triple
are minimally not rooted path graphs (see
Figure~\ref{fig:rootedSAT}).

\begin{proposition}
  The chordal graphs that are minimally not rooted path graphs and that
  contain a strong asteroidal triple are $F_1$, $F_3$, $F_4$,
  $F_5(n)_{n\geq 7}$, $F_6$, $F_7$, $F_9$, $F_{10}(n)_{n\geq 8}$,
  $F_{13}(9)$, $F_{17}(6)$.
\end{proposition}

\begin{proof}
The graphs $F_1$, $F_3$, $F_4$, $F_5(n)_{n\geq 7}$, $F_6$, $F_7$,
$F_9$,
  $F_{10}(n)_{n\geq 8}$, $F_{13}(9)$, and $F_{17}(6)$ contain a strong asteroidal
  triple, so they are forbidden in directed path graphs by
  Theorem~\ref{th:directed}, and thus forbidden in rooted path graphs. We
  omit the proof of minimality; one has to check that, for each of
  these graphs, one can remove any vertex and then one can find a
  clique rooted path tree for the resulting graph.

  Suppose there is a chordal graph $G$ that is minimally not a rooted path graph, that contains a strong asteroidal triple and that is
  different from $F_1$, $F_3$, $F_4$, $F_5(n)_{n\geq 7}$, $F_6$,
  $F_7$, $F_9$, $F_{10}(n)_{n\geq 8}$, $F_{13}(9)$, and $F_{17}(6)$.  Since $G$
  is minimally not a rooted path graph, $G$ does not strictly contain any of the graphs listed. By
  Corollary~\ref{cor:strongasteroidaltriple}, $G$ contains one of
  $F_{13}(4k+1)_{k\geq 3}$, $F_{15}(4k+2)_{k\geq 2}$,
  $F_{16}(4k+3)_{k\geq 2}$, or $F_{17}(4k+2)_{k\geq 2}$.  But all these
  graphs strictly contain $F_{23}$ of
  Figure~\ref{fig:weakasteroidalquadruple}. The graph $F_{23}$ is not
  a rooted path graph, so $G$ fails to be minimally not a rooted path graph,
a contradiction.
\end{proof}

   \begin{figure}[h]
\footnotesize
     \centering
\begin{tabular}{cccc}
\includegraphics[scale=0.5]{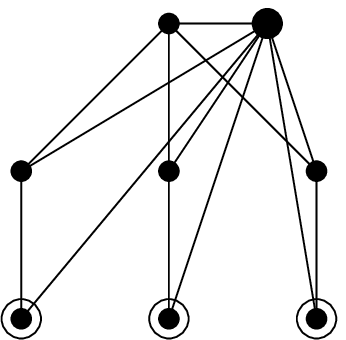} \ &\
\includegraphics[scale=0.5]{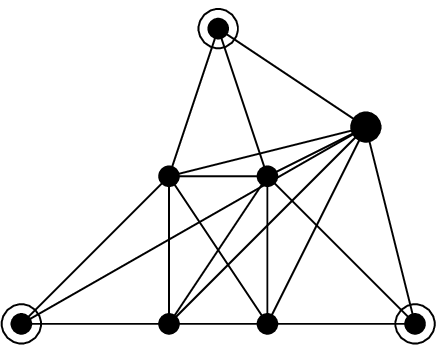} \ &\
\includegraphics[scale=0.5]{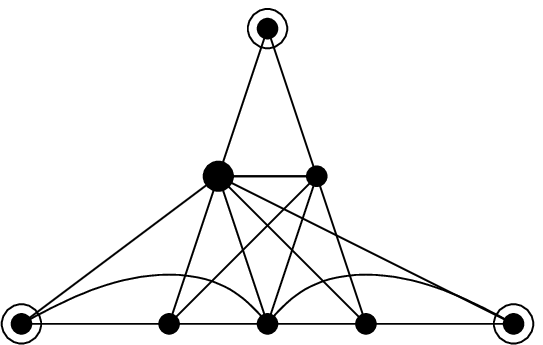} \ &\
\includegraphics[scale=0.5]{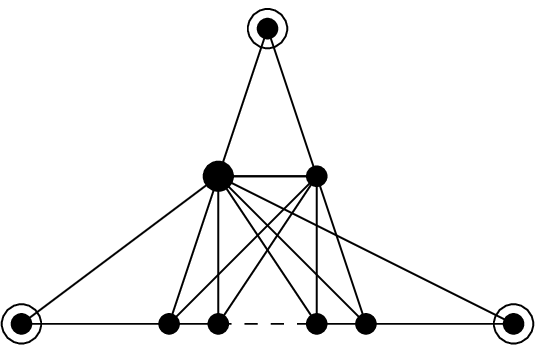} \\
$F_1$ & $F_3$ & $F_4$ & $F_5(n)_{n\geq 7}$
\end{tabular}

\begin{tabular}{cccc}
\includegraphics[scale=0.5]{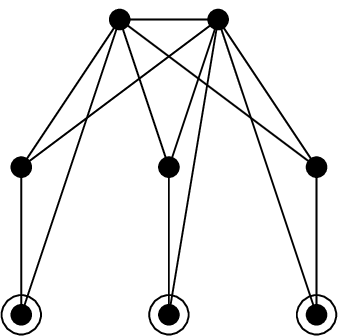} \ &\
\includegraphics[scale=0.5]{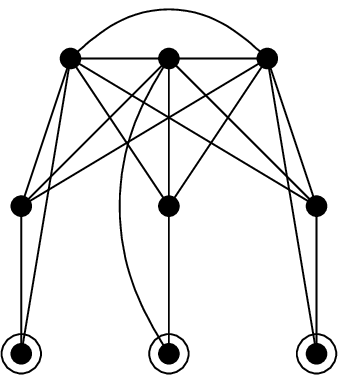} \ &\
\includegraphics[scale=0.5]{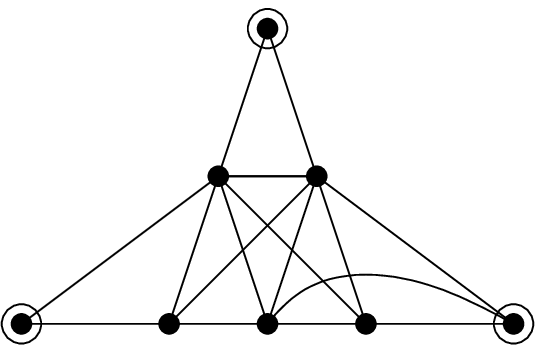}  \ &\
\includegraphics[scale=0.5]{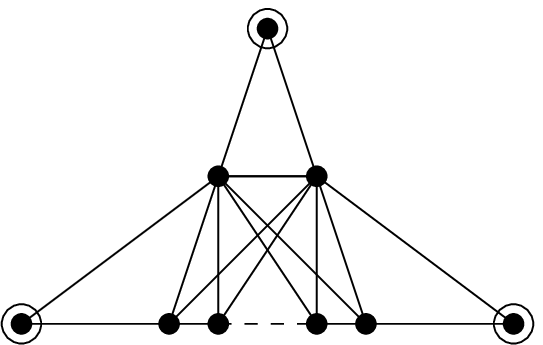} \\
$F_6$ & $F_7$ &  $F_9$ & $F_{10}(n)_{n\geq 8}$ \\
\end{tabular}

\begin{tabular}{cc}
\includegraphics[scale=0.5]{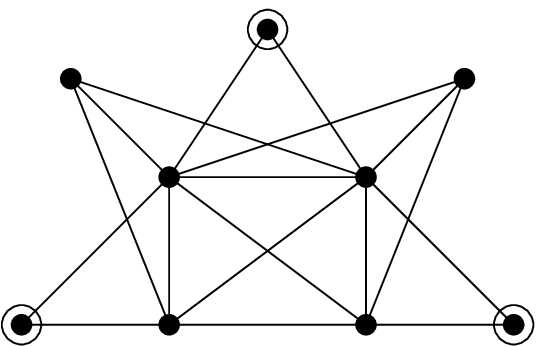}  \ &\
\includegraphics[scale=0.5]{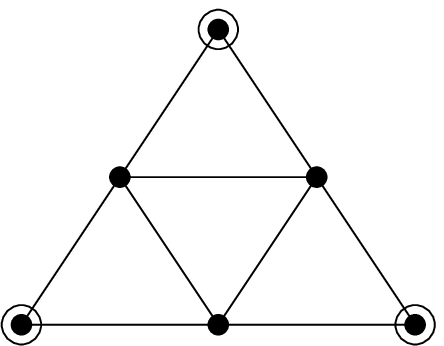} \\

 $F_{13}(9)$ &  $F_{17}(6)$  \\
\end{tabular}

\caption{Minimal forbidden induced subgraphs of rooted path graphs
  that are chordal and contain a strong asteroidal triple}
\label{fig:rootedSAT}
   \end{figure}

   Now that we know the complete list of minimal forbidden induced
   subgraphs of rooted path graphs that are chordal and contain a
   strong asteroidal triple, we can look at those that contain a weak
   asteroidal quadruple. In this case, we have not obtained a complete list.
   We need the following lemma for
   which we omit the proof.

   \begin{lemma}
     \label{lem:disjoint}
     If $G$ is a graph on four vertices such that any of its induced
     subgraphs on three vertices contains at least one edge, then $G$
     contains a triangle or two disjoint edges.
   \end{lemma}

   A \emph{parallel asteroidal quadruple} is an asteroidal quadruple
   $z_1,z_2,z_3,z_4$ such that there is a strong path linking
   $z_1$ and $z_2$ and a strong path linking $z_3$ and $z_4$. Parallel
   asteroidal quadruples are a particular type of weak asteroidal
   quadruple.  By replacing edges by strong paths in
   Lemma~\ref{lem:disjoint}, it is easy to see that a weak asteroidal
   quadruple either contains a strong asteroidal triple or is a parallel
   asteroidal quadruple. So we have the following corollary.

   \begin{corollary}
     If a chordal graph does not contain a strong asteroidal triple but does contain a
     weak asteroidal quadruple, then it contains a parallel asteroidal
     quadruple.
   \end{corollary}

The following lemma shows that we do not need to consider
attachments of type 2 when dealing with graphs that are minimally not rooted path
graphs and contain no strong asteroidal triple.

   \begin{lemma}
     \label{lem:k4}
     Let $G$ be a chordal graph that is minimally not a rooted path
     graph and that contains no strong asteroidal triple. If $u$ and
     $v$ are linked by a strong path, then they are linked by a strong
     path that uses no attachment of type 2.
   \end{lemma}

   \begin{proof}
     First, notice that the graph $F_{23}$ satisfies the hypothesis of the lemma, so we
     can assume that $G$ is different from $F_{23}$. Also $G$
     does not strictly contain $F_{23}$ as $G$ is minimally not a
     rooted path graph.  Suppose $u$ and $v$ are linked by a strong
     path. We may assume that $u$ and $v$ have no common neighbors,
     for otherwise the lemma is true. Let $X,Y,Z$ be a strong path
     linking $u$ and $v$ such that $|X\cup Y|$ is minimum. We use the
     same notation as in the definition of strong path. Suppose there
     is a clique of size four $x_i,y_j,x_{i+1},y_{j+1}$ of $X\cup Y$
     with an attachment of type 2.  We may assume that there exist
     $4t+3$ ($t\geq0$) vertices
     $z_1,\ldots,z_{2t+1},z'_1,\ldots,z'_{2t+2}$ of $Z$ such that
     vertices $x_i,y_j,x_{i+1},y_{j+1},z_1,\ldots,z_{2t+1}$ form a
     clique $Q$, vertices $z_1,\ldots,z_{2t+1}$ see exactly
     $x_i,y_j,x_{i+1},y_{j+1}$ on $X\cup Y\cup\{u,v\}$, vertices
     $z'_1,\ldots,z'_{2t+2}$ form a stable set, vertex $z'_k$ ($1\leq
     k\leq 2t+2$) sees exactly $z_{k-1},z_k$ on $Q\cup X\cup
     Y\cup\{u,v\}$ (with $z_0=x_i$ and $z_{2t+2}=x_{i+1}$).

     Let $x_0=u$ and $x_{r+1}=v$. If $x_{i-1}$ sees $y_{j+1}$, then
     there is a strong path $X',Y',Z$ between $u$ and $v$ with $X'
     \subseteq \{x_1,\ldots,x_{i-1},y_{j+1},\ldots,y_s\}, Y' \subseteq
     \{y_1,\ldots,y_{j},x_{i+1},\ldots,x_r\}$ such that $|X'\cup Y' |<
     |X\cup Y|$, a contradiction to the minimality of $|X\cup Y|$. So,
     $x_{i-1}$ misses $y_{j+1}$. Similarly, $y_{j-1}$ misses
     $x_{i+1}$. If $x_{i-1}$ misses $y_j$ and $y_{j-1}$ misses $x_i$,
     then $X\cup Y \cup\{u\}$ contains a hole.  So, there exists
     $z'_0\in\{x_{i-1},y_{j-1}\}$ that misses both $x_{i+1}$ and
     $y_{j+1}$ and sees both $x_{i}$ and $y_{j}$. Similarly, there
     exists $z'_{2t+3}\in\{x_{i+1},y_{j+1}\}$ that misses both $x_{i}$
     and $y_{j}$ and sees both $x_{i+1}$ and $y_{j+1}$.  Write
     $z_{-1}=y_{j}$, $z_0= x_{i}$, $z_{2t+2}=x_{i+1}$,
     $z_{2t+3}=y_{j+1}$.  Now, the vertices
     $z_{-1},z_0,z_1,z_2,z_3,z'_0,z'_1,z'_2,z'_3$ form the graph
     $F_{23}$, a contradiction.
   \end{proof}

   Thus, we may consider only parallel quadruples in which strong paths
   use no attachment of type 2.  Even with this restriction, one can
   obtain many minimal forbidden subgraphs of rooted path graphs, and
   we can find no simple way to represent them as a finite number of
   infinite families, as the case for interval graphs, path graphs
   or directed path graphs.

   There is no way to represent all types of strong paths, with
   attachment of type $1$, that can link two vertices in a chordal
   graph. One can put many cliques of size four between $X\cup Y$
   (each with an attachment of type $1$).  And, given such a strong
   path $X,Y,Z$ linking $z_1$ and $z_2$ and such a strong path $X',Y',Z'$
   linking $z_3$ and $z_4$ (all these vertices are distinct and there are no edges between
   the two parts), one can easily construct a graph which is minimally not a
   rooted path graph by adding a path $v_1$-$\cdots$-$v_\ell$
   ($\ell\geq 1$) where $v_1$ sees all the vertices of $X\cup Y \cup Z$
   and $v_\ell$ sees all the vertices of $X'\cup Y' \cup Z'$ (see
   Figure~\ref{fig:waqinfinite}).

\begin{figure}[h]
   \centering
\includegraphics[scale=0.5]{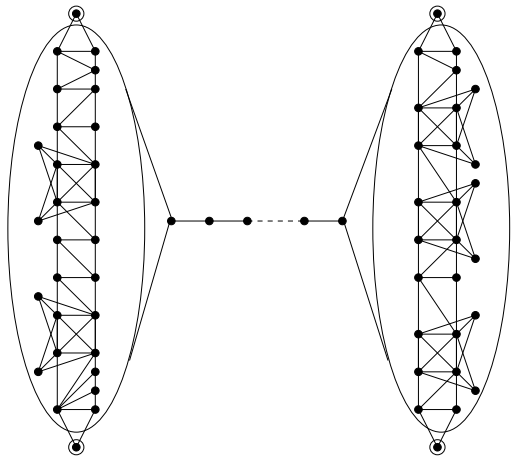}
\caption{Example of graph which is minimally not a rooted path graph and which contains a
  weak asteroidal quadruple}
   \label{fig:waqinfinite}
 \end{figure}

 And, there are many more different possibilities as the two strong
 paths may share some vertices, they may have some edges between them,
 the attachment of type $1$ can see more than $3$ vertices on the strong
 path, they can be used in more than one clique of size four, etc.  In
 Figure~\ref{fig:DPG}, we give more examples of graphs which are minimally not
 rooted path graphs and which contain a weak asteroidal quadruple where
 strong paths are just a common neighbor. Even with this restriction,
 they are many variants and we give just a few.

\begin{figure}[h]
\centering

\begin{tabular}{cc}
\includegraphics[scale=0.5]{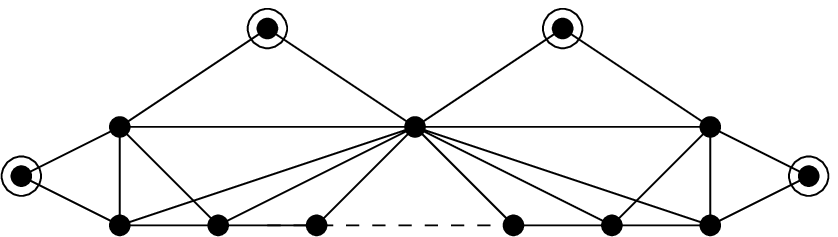} \ &\
\includegraphics[scale=0.5]{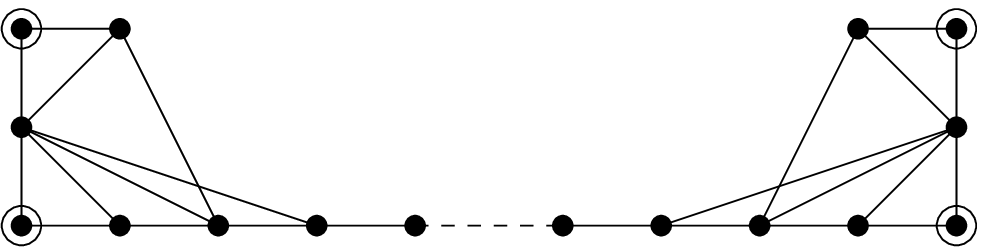} \\
\includegraphics[scale=0.5]{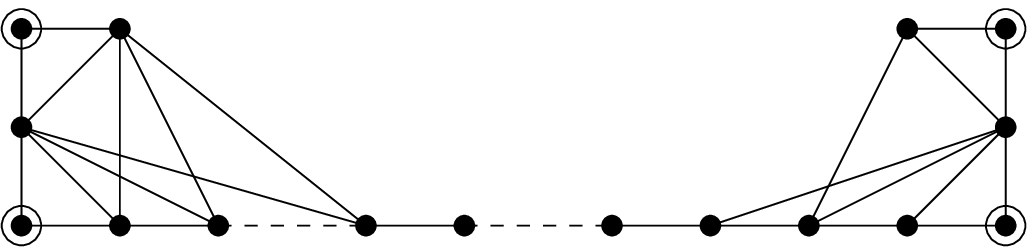} \ &\
\includegraphics[scale=0.5]{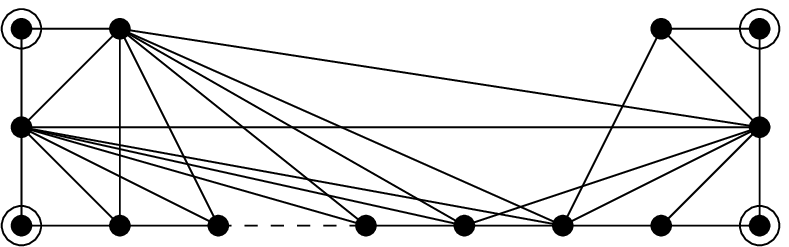} \\ \\
\includegraphics[scale=0.5]{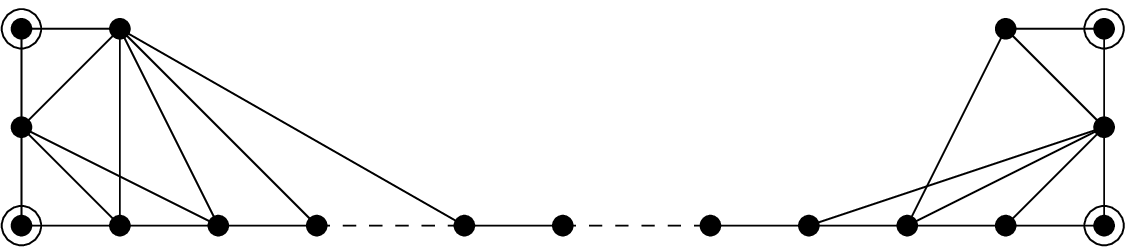} \ &\
\includegraphics[scale=0.5]{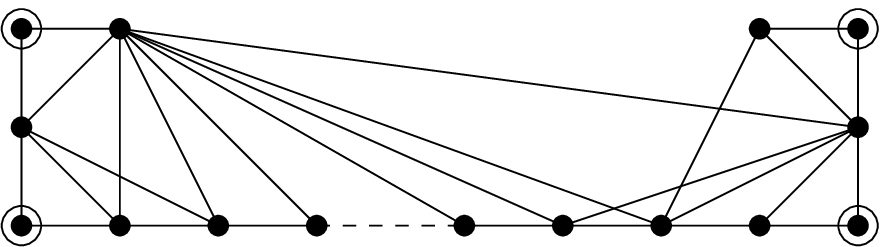} \\ \\
\includegraphics[scale=0.5]{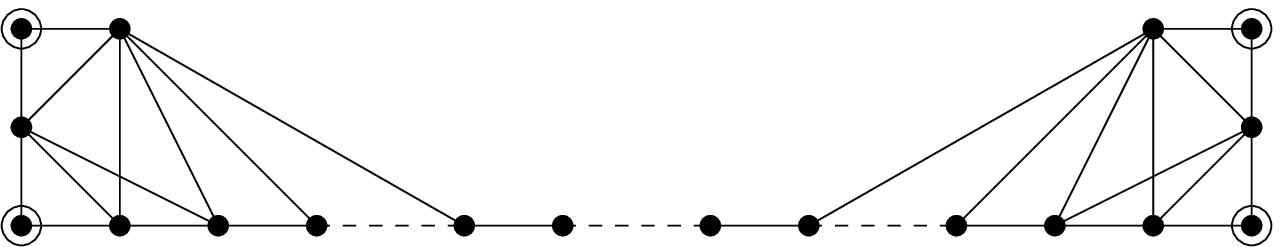} \ &\
\includegraphics[scale=0.5]{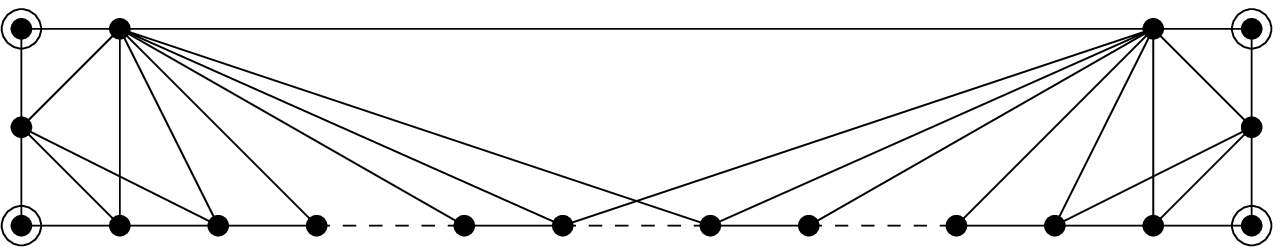} \\ \\
\end{tabular}

\caption{Examples of minimal forbidden induced subgraphs of rooted path
  graphs which contain a weak asteroidal quadruple}
\label{fig:DPG}
    \end{figure}

\section{Conclusion}\label{sec:conclusion}

We have defined particular asteroids on three and four vertices to
obtain a characterization of directed path graphs and some partial
results on rooted path graphs. The characterization of rooted path
graphs that we conjecture will be a nice alternative to finding
the list of all minimal forbidden induced subgraphs for this
class.

One can also try to do similar work for path graphs. Path graphs
are a superclass of directed path graphs that may contain some
strong asteroidal triples (odd suns). Can one define a particular
type of strong asteroidal triple that will give a nice
characterization of path graphs?

   \begin{figure}[h!]
\footnotesize     \centering
\begin{tabular}{c}
\includegraphics[scale=0.35]{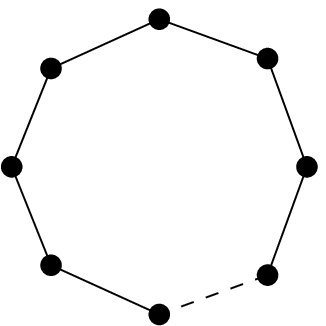} \\
$F_{0}(n)_{n\geq 4}$ \\
\end{tabular}

\begin{tabular}{ccccc}
\includegraphics[scale=0.35]{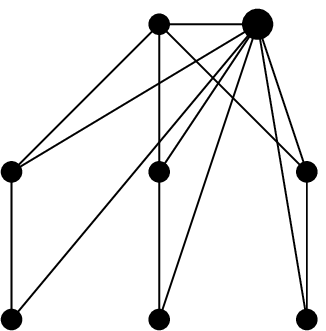} \ &\
\includegraphics[scale=0.35]{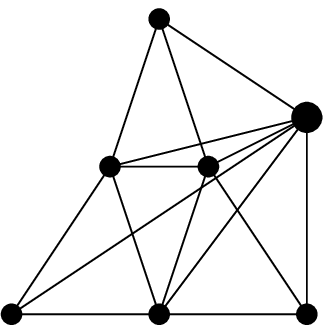} \ &\
\includegraphics[scale=0.35]{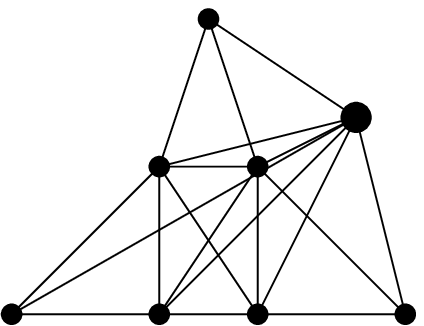} \ &\
\includegraphics[scale=0.35]{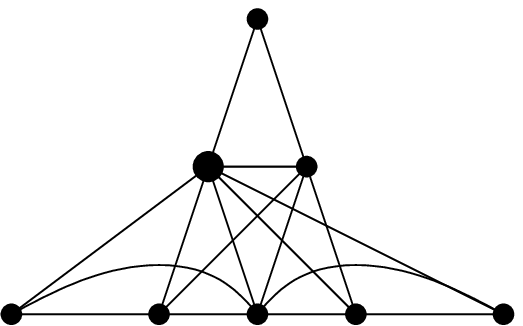} \ &\
\includegraphics[scale=0.35]{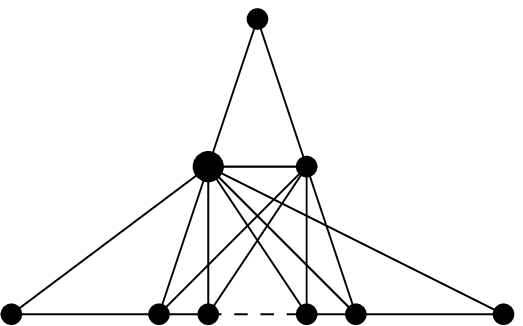} \\
$F_1$ & $F_2$ & $F_3$ & $F_4$ & $F_5(n)_{n\geq 7}$
\end{tabular}

\begin{tabular}{ccccc}
\includegraphics[scale=0.35]{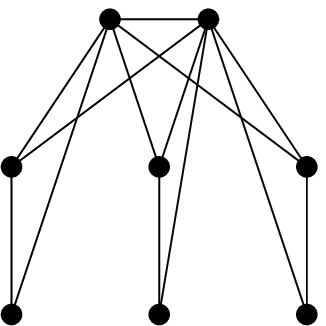} \ &\
\includegraphics[scale=0.35]{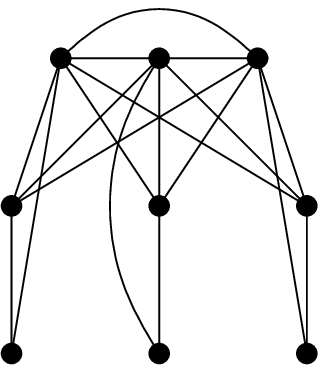} \ &\
\includegraphics[scale=0.35]{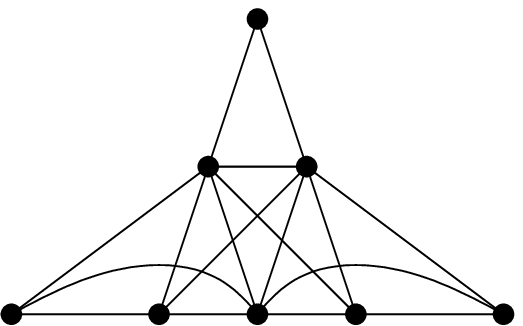} \ &\
\includegraphics[scale=0.35]{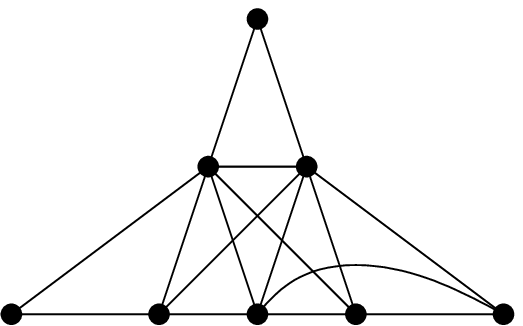}  \ &\
\includegraphics[scale=0.35]{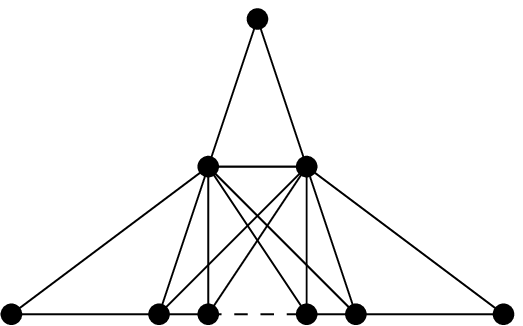} \\
$F_6$ & $F_7$ & $F_8$ & $F_9$ & $F_{10}(n)_{n\geq 8}$ \\
\end{tabular}

\begin{tabular}{ccccccc}
\includegraphics[scale=0.35]{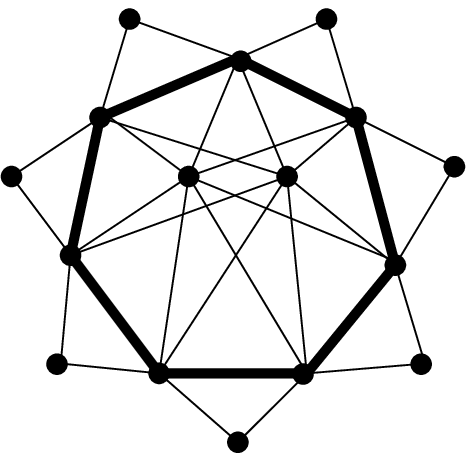} \ &\
\includegraphics[scale=0.35]{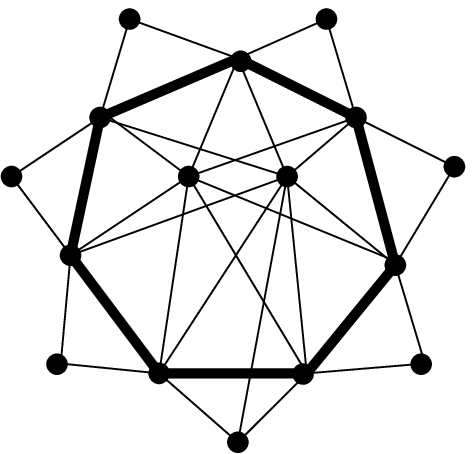} \ &\
\includegraphics[scale=0.35]{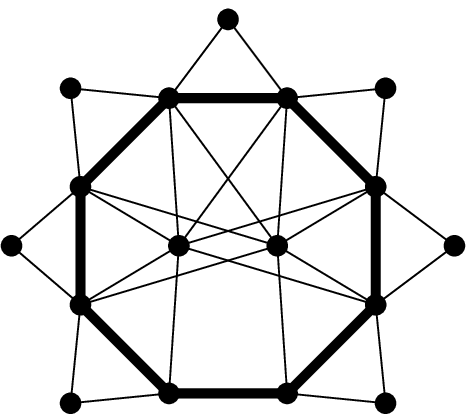}\ &\
\includegraphics[scale=0.35]{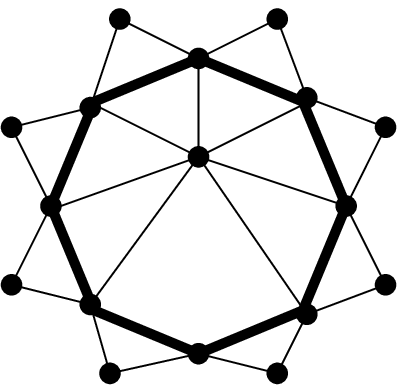}  \ &\
\includegraphics[scale=0.35]{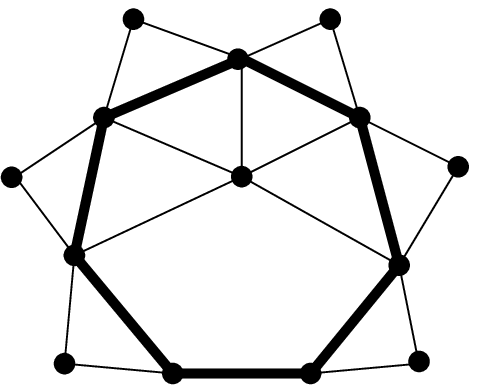} \ &\
\includegraphics[scale=0.35]{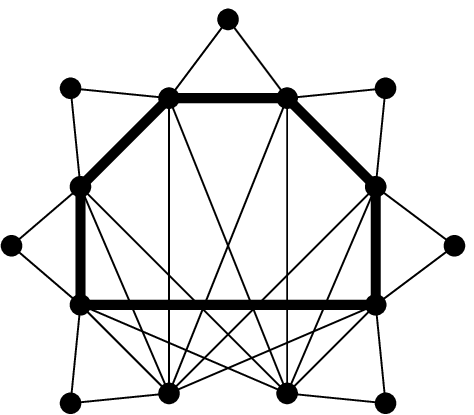} \\
$F_{11}(4k)_{k\geq 2}$ & $F_{12}(4k)_{k\geq 2}$ & $F_{13}(4k+1)_{k\geq 2}$ &
 $F_{14}(4k+1)_{k\geq 2}$ & $F_{15}(4k+2)_{k\geq 2}$ &
$F_{16}(4k+3)_{k\geq 2}$ \\
\end{tabular}

\caption{ Minimal forbidden induced subgraphs for path graphs (bold edges form a clique)}
\label{fig:path}
   \end{figure}
   \begin{figure}[h!]
\footnotesize
     \centering
\begin{tabular}{c}
\includegraphics[scale=0.35]{zfig-f0.eps} \\
$F_0(n)_{n\geq 4}$ \\
\end{tabular}

\begin{tabular}{cccc}
\includegraphics[scale=0.35]{zfig-f1-r.eps} \ &\
\includegraphics[scale=0.35]{zfig-f3-r.eps} \ &\
\includegraphics[scale=0.35]{zfig-f4-r.eps} \ &\
\includegraphics[scale=0.35]{zfig-f5-r.eps} \\
$F_1$ & $F_3$ & $F_4$ & $F_5(n)_{n\geq 7}$
\end{tabular}

\begin{tabular}{cccc}
\includegraphics[scale=0.35]{zfig-f6-r.eps} \ &\
\includegraphics[scale=0.35]{zfig-f7-r.eps} \ &\
\includegraphics[scale=0.35]{zfig-f9-r.eps}  \ &\
\includegraphics[scale=0.35]{zfig-f10-r.eps} \\
$F_6$ & $F_7$ &  $F_9$ & $F_{10}(n)_{n\geq 8}$ \\
\end{tabular}

\begin{tabular}{ccccccc}
\includegraphics[scale=0.35]{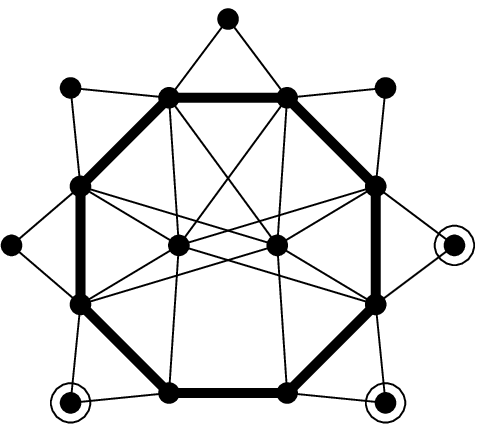}\ &\
\includegraphics[scale=0.35]{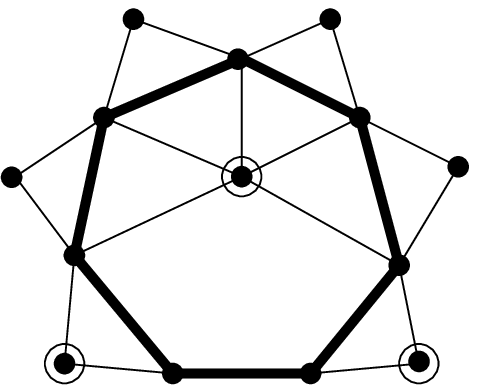} \ &\
\includegraphics[scale=0.35]{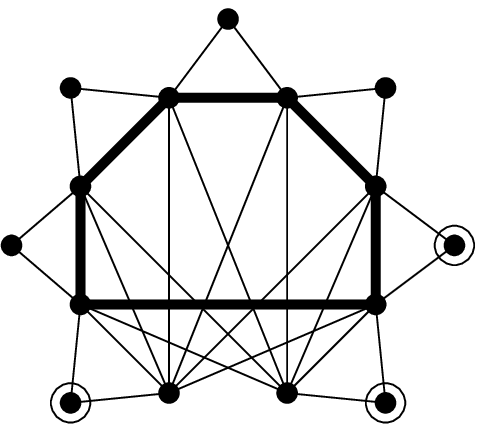} \ &\
\includegraphics[scale=0.35]{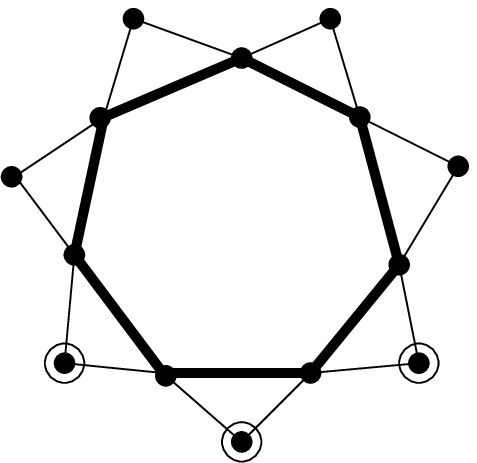} \\
 $F_{13}(4k+1)_{k\geq 2}$ &
$F_{15}(4k+2)_{k\geq 2}$ &
$F_{16}(4k+3)_{k\geq 2}$ & $F_{17}(4k+2)_{k\geq 1}$  \\
\end{tabular}

\caption{ Minimal forbidden induced subgraphs for directed graphs (bold edges form a clique)}
\label{fig:directed}
   \end{figure}
   \begin{figure}[h!]
\footnotesize
     \centering
\begin{tabular}{ccccc}
\includegraphics[scale=0.35]{zfig-f0.eps} \ &\
\includegraphics[scale=0.35]{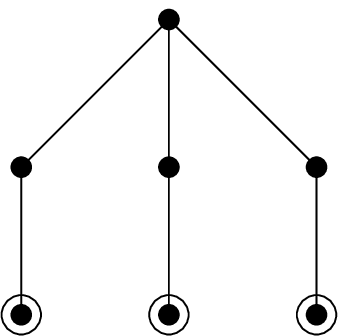} \ &\
\includegraphics[scale=0.35]{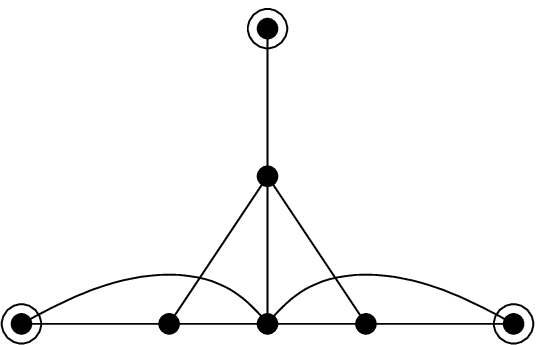} \ &\
\includegraphics[scale=0.35]{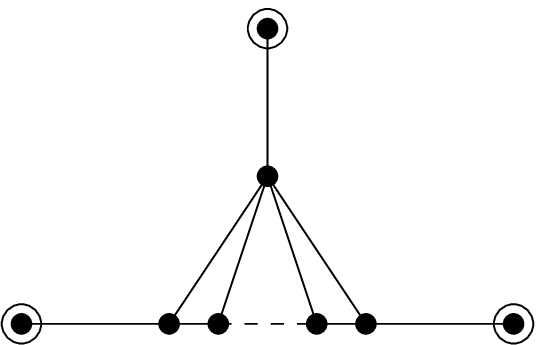} \ &\
\includegraphics[scale=0.35]{zfig-f10-r.eps} \\
$F_{0}(n)_{n\geq 4}$ & $F_{18}$ & $F_{19}$ & $F_{20}(n)_{n\geq 6}$ &
$F_{21}(n)_{n\geq 6}$
\end{tabular}

\caption{ Minimal forbidden induced subgraphs for interval graphs}
\label{fig:interval}
   \end{figure}

\end{document}